\documentclass[prb, aps, amsmath, amssymb, twocolumn, showpacs, superscriptaddress]{revtex4-2}

\usepackage{dcolumn}
\usepackage{bm}
\usepackage{physics}
\usepackage{xcolor}

\usepackage{graphicx}
\usepackage[normalem]{ulem}
\usepackage[colorlinks=true,linkcolor=blue,citecolor=blue]{hyperref}

\begin{document}


\title{
    On the Entanglement Entropy Distribution of a Hybrid Quantum Circuit 
}

\author{Jeonghyeok Park}
\affiliation{NextQuantum Innovation Research Center, Department of Physics and Astronomy, Seoul National University, 08826 Seoul, South Korea}
\author{Hyukjoon Kwon}%
\email{hjkwon@kias.re.kr}
\affiliation{%
 School of Computational Sciences, Korea Institute for Advanced Study, Seoul 02455, South Korea
}%

\author{Hyunseok Jeong}
\email{jeongh@snu.ac.kr}
\affiliation{Department of Physics and Astronomy, Seoul National University, 08826 Seoul, South Korea}

\begin{abstract}
We investigate the distribution of entanglement entropy in hybrid quantum circuits consisting of random unitary gates and local measurements applied at a finite rate. We demonstrate that higher moments of the entanglement entropy distribution, such as the ratio of the variance to the mean and the skewness, capture nontrivial features of the measurement-induced dynamics that are invisible to the mean entropy alone. We demonstrate that these quantities exhibit distinct and robust behaviors across the volume-law and area-law phases, and can serve as effective diagnostics of measurement-induced entanglement transitions. We propose a phenomenological model describing the effect of measurements in the area-law regime, which, when combined with the directed polymer in a random environment description of the volume-law phase, matches numerical simulations well across both phases.
\end{abstract}

\maketitle


\section{\label{sec:level1}Introduction}
Random unitary circuits (RUCs) provide a simple paradigmatic model that extends our understanding of the chaotic evolution of quantum many-body systems~\cite{RandomQuantumCircuits, PhysRevLett.98.130502, Hosur2016,5d6p-8d1b}. Accompanied by chaotic scrambling of information, random unitary evolution results in an extensive scaling of the entanglement entropy of the system~\cite{PhysRevX.7.031016, PhysRevX.8.021014, PhysRevB.99.174205}. Contrary to unitary evolution, measurement on a quantum system is non-unitary and irreversible. The peculiar properties of measurement give rise to rich phenomena in various quantum systems~\cite{wiseman2010quantum}. In quantum information science, measurement-based quantum computation utilizes such properties \cite{PhysRevLett.86.5188,briegel_measurement-based_2009}. In many-body physics, the quantum Zeno effect demonstrates that frequent measurements result in an area-law scaling of the entanglement entropy of the system~\cite{10.1063/1.523304,PhysRevB.98.205136}.

Hybrid quantum circuits, which consist of unitary operators and measurements, result in stochastic trajectories of quantum states. The stochastic nature of the states arises from multiple factors of randomness: randomly sampled unitary operators, measurement locations, and measurement outcomes with probabilities given by Born's rule that determine the post-measurement states. Different measurement outcomes give rise to different trajectories of quantum states with entanglement entropy that fluctuates about a mean value. Depending on how frequently the measurement operators act on qubits, the mean entanglement entropy exhibits two different scaling behaviors as the measurement rate varies. From the extensive, i.e., volume-law, scaling of the entanglement entropy to the area-law scaling of the entropy, a measurement-induced phase transition (MIPT) occurs between these two phases as the measurement rate of the hybrid circuit crosses the critical probability, $p_{c}$~\cite{PhysRevX.9.031009,PhysRevB.100.134306}.

\begin{figure}[t]
    \centering
    \includegraphics[width=\linewidth]{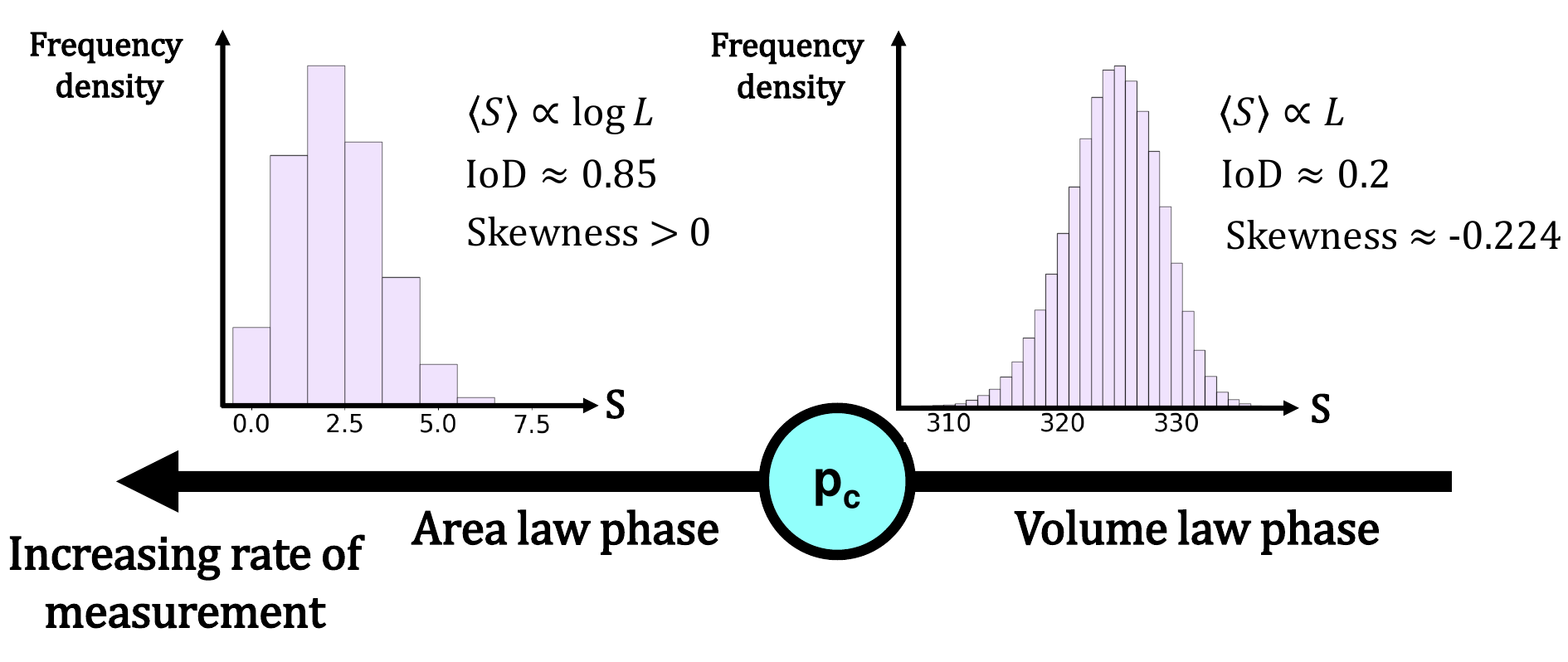}
    \caption{Schematic diagram of the entanglement entropy spectrum of hybrid circuits as the measurement rate increases. The changing shape can be quantified by the skewness. Each value of the index of dispersion (IoD) corresponds to each of the distributions shown.}
    \label{fig:scheme}
\end{figure}

MIPT has been extensively studied in various fields of physics.
Since the first proposal~\cite{PhysRevB.98.205136} consisting of random unitary circuits with local measurements~\cite{PhysRevB.98.205136,PhysRevX.9.031009,PhysRevB.100.134306,jian_measurement_induced_2020,PRXQuantum.2.010352,RandomQuantumCircuits, PhysRevB.100.064204, PhysRevX.11.011030, PRXQuantum.2.010352}, such a formalism has been extended to a wide range of many-body systems such as free fermionic systems~\cite{PhysRevX.13.041046,PhysRevLett.132.110403, 3zfd-3hqt}, Ising-like models~\cite{ising1,ising2}, and bosonic systems~\cite{y5r3tv78}. In the context of quantum information science, MIPT has been extended to information exchange processes~\cite{PhysRevA.111.L010402}. In understanding the entanglement dynamics of the hybrid circuit, the non-unitarity of the measurements poses a significant challenge to the analysis of the transition. While mappings of hybrid quantum circuits to statistical-mechanics models have successfully yielded scaling properties~\cite{PhysRevB.101.104301, jian_measurement_induced_2020} and provided an exact mapping in the volume-law phase~\cite{PRXQuantum.4.010331}, a complete understanding of the entanglement behavior across the entire range of measurement probabilities remains open, especially in the area-law phase. 

\begin{figure*}
    \centering
    \includegraphics[width=0.98\textwidth]{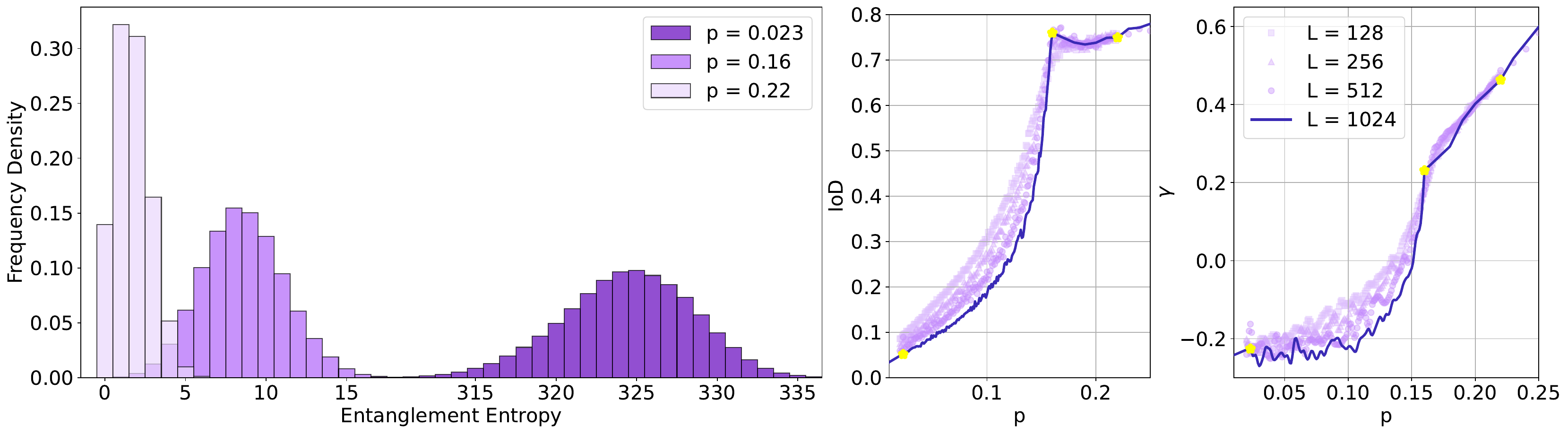}

    \caption{(Left) Histograms of entanglement entropy distributions of the hybrid circuits at measurement rates $p = 0.023, 0.16$, and $0.22$, which lie in the volume-law phase, at criticality, and in the area-law phase, respectively. (Middle) IoD curve and (Right) skewness curve with increasing rate of measurement, $p$, with yellow markers indicating values of $p$ that correspond to the histograms. }
        \label{fig:dist_IoD_skew}
\end{figure*}


The MIPT is usually characterized by the scaling of the mean entanglement entropy, with only a few works examining higher moments such as variance~\cite{PhysRevB.100.064204, NatPhy.19.1314-1319}. However, hybrid quantum circuits are intrinsically non-self-averaging due to the stochastic nature of measurements. As a result, the entanglement entropy exhibits strong trajectory-to-trajectory fluctuations that persist even in the thermodynamic limit~\cite{PhysRevLett.77.3700,PhysRevB.101.174312}. 
Moreover, in a directed polymer in a random environment (DPRE) \cite{PRXQuantum.4.010331}, which provides an effective description of hybrid circuits in the volume-law phase, the entanglement entropy plays the role of a fluctuating free energy, whose higher moments encode universal features of the underlying stochastic dynamics~\cite{PhysRevE.52.3469}. In such a situation, the average entanglement entropy captures only the leading extensive contribution, while finer statistical structures could be hidden in higher-order moments.

In this work, we explore higher-order moments of the entanglement entropy distributions in hybrid quantum circuits beyond their mean values. We first introduce the index of dispersion (IoD), defined as the ratio of the variance to the mean, and demonstrate that it exhibits qualitatively distinct behaviors in the volume-law and area-law phases. We further observe an abrupt change in the IoD near the critical point, enabling a precise estimate of $p_c$, which cannot be obtained from the variance alone. We further analyze a higher-order moment of the entanglement entropy distribution by examining its skewness, which quantifies the distribution’s asymmetry using moments up to third order. Remarkably, independent of the system size, the skewness remains constant in the volume law regime, while it follows power-law scaling in the area-law regime. Similar to the IoD, the skewness increases sharply near the critical probability, providing a sensitive and robust diagnostic of MIPT in the hybrid circuit.

We also explore effective descriptions of the entanglement entropy distribution in both phases of the hybrid circuit. In the volume-law regime, we analyze the higher-order moments of the entanglement distribution using the established DPRE model~\cite{PRXQuantum.4.010331}, yielding precise agreement with the simulation results. In the area-law regime, we introduce a minimal coarse-grained stochastic model based on a Bell pair description of entanglement in the stabilizer formalism~\cite{Ent_stab}. We compare the distributions predicted by these models with our numerical results and evaluate the statistical distance between them using the Kullback–Leibler divergence.

Our results demonstrate that there exists a rich structure of the entanglement behavior by exploring their distribution beyond the mean values. This provides a novel insight into understanding MIPT at the statistical level, where the proposed methodology can also be applied to other stochastic quantum circuits, such as hybrid quantum circuits with conservation laws~\cite{PhysRevX.12.041002,PhysRevB.110.L140301}, including noise~\cite{PhysRevLett.132.240402,PhysRevB.110.064323}, or magic as the quantity of interest~\cite{PRXQuantum.5.030332,Niroula2024}.

\section{Hybrid quantum circuit}
\subsection{Model description}
We consider a one-dimensional quantum circuit with a brickwork arrangement of random unitary gates and measurement operators acting on each qubit at a rate of $p$. Such a hybrid quantum circuit has been widely studied in the context of measurement-induced phase transition (MIPT) \cite{PhysRevB.100.134306, PhysRevX.9.031009}. Here, we explain the details of the model for the sake of completeness.

\begin{figure}
    \centering
    \includegraphics[width=0.9\linewidth]{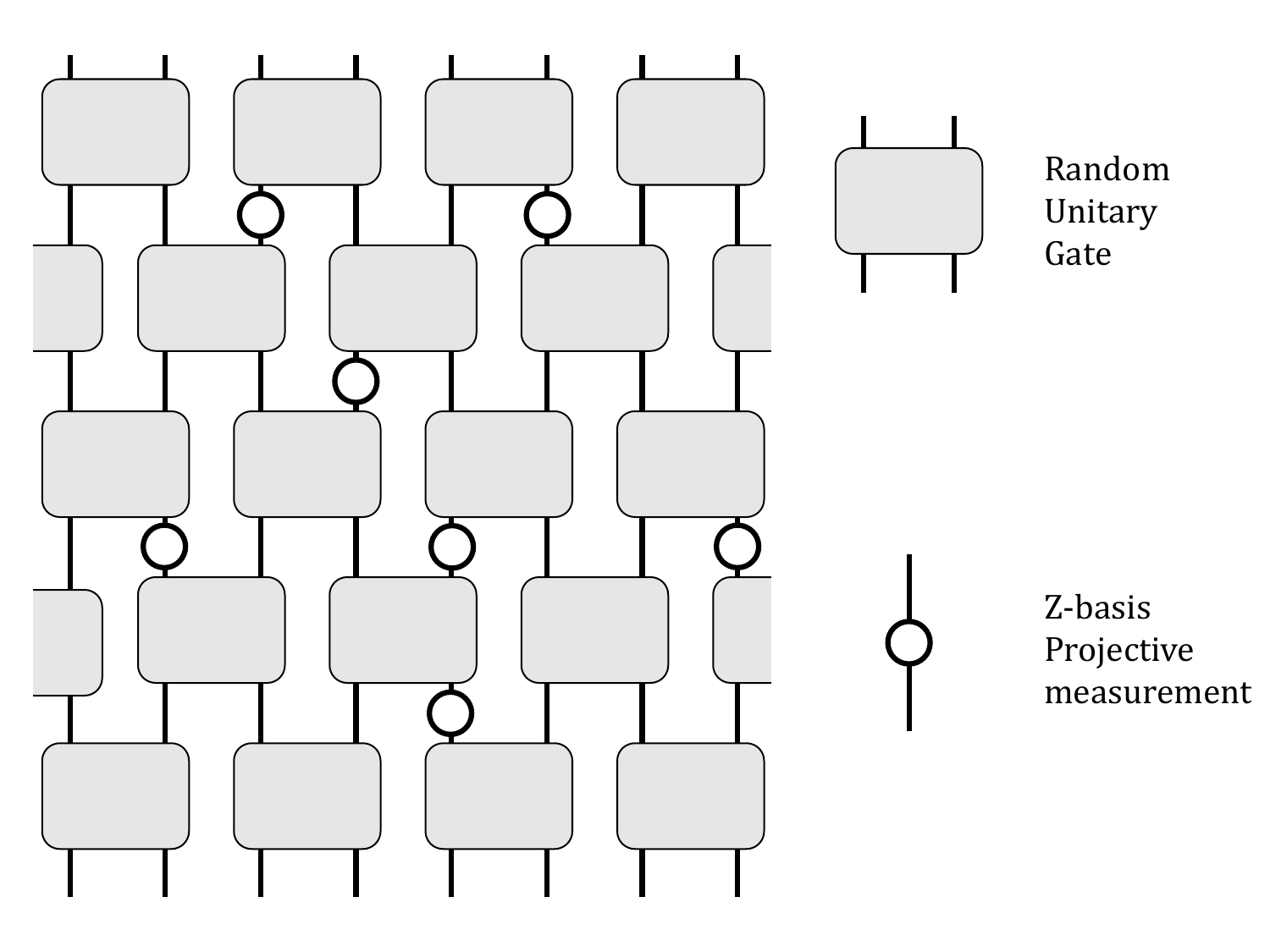}
    \caption{A hybrid quantum circuit consisting of random unitary gates sampled from uniformly distributed two qubit Clifford gates arranged in a brickwork fashion and projective measurements acting on each qubit at a rate $p$.}
    \label{fig:circuit}
\end{figure}

In our model (see Fig.~\ref{fig:circuit}), a quantum system consists of $L$ qubits initialized in a product state, $\ket{\Psi(t=0)} = \ket{0}^{\otimes L}$. While we choose the computational basis state $\ket{0}^{\otimes L}$ for convenience, the choice of initial state does not affect the entanglement dynamics after reaching the steady state. In each discrete time step, we apply layers of random Clifford unitary gates on the system in an alternating fashion,
\begin{align}
    \ket{\Psi} &\rightarrow U_{L-1,L}\cdots U_{2,3}U_{0,1} \ket{\Psi} &&\mathrm{(even~time~step)} \\
    \ket{\Psi} &\rightarrow U_{L,0}\cdots U_{3,4} U_{1,2} \ket{\Psi}  &&\mathrm{(odd~time~step)}.
\end{align}
Each $U_{i,i+1}$ acts on a neighboring pair of qubits and is sampled independently from a uniform distribution over two-qubit Clifford operators \cite{Hadamard_Free}, which provides an efficient approximation to Haar-random unitaries while remaining computationally tractable \cite{webb2016clifford,PhysRevA.70.052328}. In terms of a quantum circuit, the evolution is represented as a one-dimensional brickwork circuit of random unitary gates. We also take periodic boundary conditions such that the odd layers connect the first and last qubits.

The quantity of interest is the bipartite entanglement entropy of the system:
\begin{equation}
    S = -\Tr \left[ \rho_A \log \rho_A \right],
\end{equation}
where $\rho_A = \Tr_B \ket{\Psi}\bra{\Psi}$. The bipartition between $A$ and $B$ is drawn in the middle of the chain such that each subsystem contains the same number of contiguous qubits. In the absence of measurements, the alternating structure of the unitary layers efficiently scrambles information across the bipartition, ultimately resulting in a maximally entangled steady state.

Including measurement operators counteracts the growth of entanglement by destroying quantum correlations between the subsystems. Once a projective measurement acts on a qubit, the resulting state is determined probabilistically as
\begin{equation}
    \ket{\Psi}_{m} = \frac{P_m \ket{\Psi}}{ \sqrt{p_m} },
\end{equation}
where $m\in \{ 0,1\}$ denotes the measurement outcome in the computational basis, $P_m$ is a projector onto the states corresponding to $m$, and the probability of observing $m$ is given by Born's rule, $p(m) = \Tr ( \ket{\Psi} \bra{\Psi} P_m )$. 
A measurement acting on the $j^{\mathrm{th}}$ qubit disentangles the measured qubit from the rest of the system.
The measurement is incorporated into the state evolution by probabilistically measuring each qubit at each time step. Therefore, the mean number of measured qubits in each time step is equal to $p L$ where $p$ is the rate of measurement acting on the qubits.

The combined action of random unitary gates and measurements, which increase and decrease entanglement, respectively, generates stochastic quantum trajectories. 
After a transient period, the circuit reaches a steady state, with the mean entanglement entropy depending on the measurement rate $p$. For different realizations, the entanglement entropy can fluctuate around the mean value. Hence, from the ensemble of trajectories for a given measurement rate $p$, we obtain the full distribution of entanglement entropy.

\begin{figure}[t]
    \centering
    \includegraphics[width=.9\linewidth]{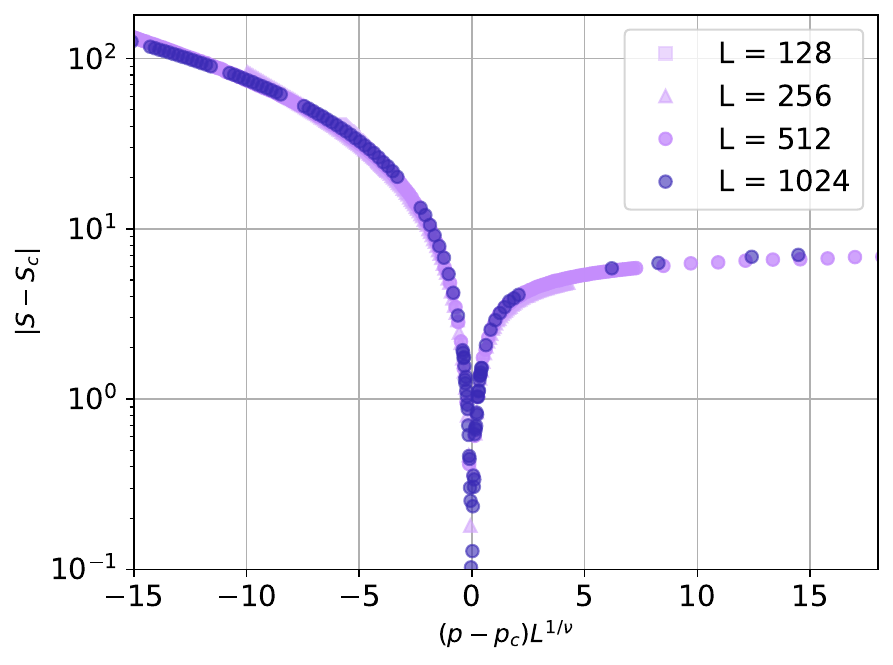}
    \caption{Collapse of the entanglement entropy of different system sizes according to the scaling form in Eq.~\eqref{eq:Scaling}. We reaffirm the known critical point, $p_c = 0.16$ and $\nu = 1.3$.}
    \label{fig:finite_size_scaling}
\end{figure}

\subsection{Average entanglement and MIPT}
Depending on the rate of measurement acting on the qubits, we observe a transition in the scaling laws of the average entanglement entropy of the system, the so-called measurement-induced phase transition (MIPT)~\cite{PhysRevX.9.031009,PhysRevB.100.134306, PhysRevX.9.031009, PhysRevB.100.064204,jian_measurement_induced_2020, PhysRevB.101.104301}. Here, we describe the behavior of the average entanglement entropy, $\langle S \rangle$, where we define $\langle \cdot \rangle$ as the average over all possible trajectories. In the earliest works in MIPT~\cite{PhysRevX.9.031009, PhysRevB.100.134306}, data collapse after finite-size scaling was adopted to locate the critical rate of measurement, $p_c$, across which the average entanglement exhibits two distinct behaviors.

In the regime where the rate of measurement is sufficiently low, i.e., $p<p_c$, quantum correlations in the system prevail, so that the average entanglement entropy of the system follows the volume law scaling, $\langle S \rangle \propto L$. On the other hand, frequent measurements on the system, i.e., $p>p_c$, prevents correlations among the qubits from
forming, resulting in the area-law scaling of the entanglement entropy, $\langle S \rangle \propto L^0$. 

Between these two scaling phases, there exists a phase transition where the entanglement entropy scales logarithmically as $\langle S \rangle \propto \log L$. Inspired by the classical statistical mechanics model~\cite{PhysRevB.100.134203,PhysRevX.9.031009, PhysRevB.100.134306}, we can use finite-size scaling to locate the critical probability and the critical exponent, $\nu$, of the phase transition:
\begin{equation} \label{eq:Scaling}
    \abs{\langle S(p)\rangle - \langle S(p_c)\rangle} = \Tilde{F} [(p-p_c)L^{1/\nu}],
\end{equation}
where $\Tilde F$ is a scaling function. The points of $\langle S\rangle$ of different system sizes collapse onto the same curve. By minimizing least squares on the collapsed curve~\cite{PhysRevX.9.031009}, we reaffirm the known value for the critical point, $p_c = 0.16$ and $\nu = 1.3$ \cite{PhysRevB.100.134306}. In Fig.~\ref{fig:finite_size_scaling}, we show data collapsing onto the same curve with the critical point value we reaffirmed.

We note that a hybrid quantum circuit with random unitary gates sampled from the Haar measure has a different critical point. In Ref.~\cite{PhysRevX.9.031009}, $p_c = 0.26$  and $\nu = 2.01$. However, the scaling laws of the average entanglement entropy are the same as in the Clifford case~\cite{PhysRevX.9.031009, PhysRevB.100.134306}.

\section{Higher-order moments of the entanglement distribution}

\begin{figure}[t]
    \centering
    \includegraphics[width=\linewidth]{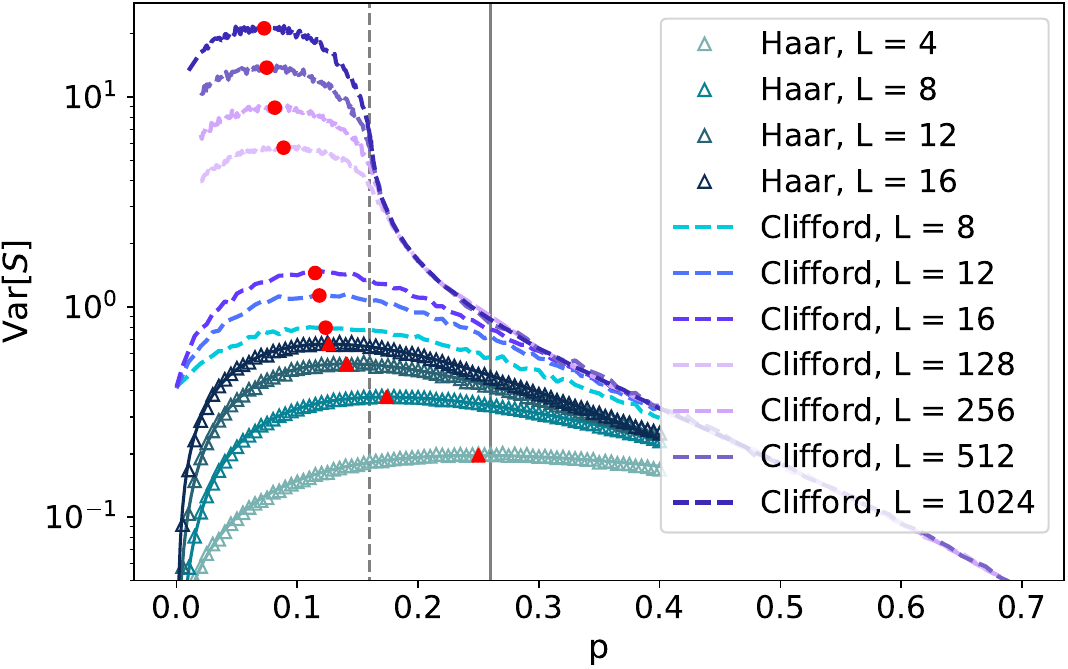}
    \caption{Variance of Haar random and Clifford random hybrid quantum circuits. The triangle (circle) points indicate the peak of the variance for the Haar (Clifford) random circuit.  The critical measurement rate for the Haar (Clifford) random circuit is indicated by a solid (dashed) vertical line. $p_c = 0.26$ for Haar random circuit and $p_c = 0.16$ for Clifford random circuit.}
    \label{fig:var_comb}
\end{figure}

\subsection{Variance}
In many phase transitions, fluctuations of a suitable diagnostic observable near criticality often sharply diverge in the thermodynamic limit, while manifesting as a peak in finite-size systems~\cite{LandauLifshitzStatPhys1, PhysRevB.99.104205}. This motivates examining the variance of the entanglement entropy,
\begin{equation}
    \mathrm{Var}[S] =\langle (S-\langle S\rangle )^2\rangle,
\end{equation}
as a potential indicator of the MIPT. Indeed, previous works~\cite{PhysRevB.100.064204, NatPhy.19.1314-1319} proposed a method to identify the critical measurement rate by locating the point at which the variance of the entanglement entropy is maximal based on phenomenological observations. It should be noted that our model with projective measurements differs from the model in the Refs.~\cite{PhysRevB.100.064204, NatPhy.19.1314-1319} with measurements of variable measurement strength. In the limit of infinite measurement strength, the variable-strength measurement converges to the projective measurement~\cite{PhysRevA.63.062305,Jacobs01092006} and the two models would coincide in such a limit. Based on such convergence, one may anticipate the validity of the method in hybrid quantum circuits with projective measurements. However, as we will discuss in the next paragraph, variance alone is insufficient in separating the phases in the system we consider.

While observing a peak of the variance can partially indicate the critical point for both Haar and Clifford random circuits for a small system size $L$ (see Fig.~\ref{fig:var_comb}), our numerical results for Clifford random circuits with a larger system size $L$ indicate that the variance might not be a reliable diagnostic of the MIPT (see Fig.~\ref{fig:var_comb}).

In particular, the peak of the variance for large system sizes deviates from $p_c$ and lies within the volume law phase. Moreover, the variance exhibits a sharp peak deep within the volume-law regime, rendering the precise location of the maximum ambiguous. These observations support the conclusion that variance alone may not provide sufficient evidence for identifying the critical point of the hybrid quantum circuit.

The discrepancy between the peak of the variance $p_{\max}$ and the critical point $p_c$ can be schematically understood from the entanglement entropy distributions. As the measurement rate increases, the mean value of entanglement entropy becomes smaller, and since the entanglement entropy cannot be below zero, the distribution of entanglement entropy becomes concentrated around $\langle S \rangle$ with lower variance. At the critical point, the mean entanglement scales logarithmically, i.e., $\langle S\rangle \propto \log L$, for which the variance is also typically bounded by the same scaling. Whereas, in the volume law regime, the distribution with sufficiently large mean is not affected severely by the $S\geq 0$ condition, and hence we see the maximal variance in the volume law regime.

\subsection{Index of dispersion}
The observation that variance alone may be insufficient to fully characterize MIPT motivates us to develop a new measure that more sensitively reflects changes in the entanglement distribution with respect to the measurement rate. As an illustrative example, when the entanglement entropy is bounded by a condition $S \geq 0$, the variance of the entanglement distribution in the area-law regime is bounded by $\mathrm{Var} [S] \sim L^0$. This implies that considering the variance rescaled by the mean value would be more appropriate in accounting for the change in entanglement distribution.

To this end, we introduce the index of dispersion (IoD), defined as the ratio of the variance of the entanglement entropy distribution to its mean,
\begin{equation} \label{eq:IOD}
    \mathrm{IoD} = \frac{ \mathrm{Var}[S] }{\langle S\rangle}.
\end{equation}
For a Poisson distribution, $\mathrm{IoD} = 1$, while for a binomial distribution, $0<\mathrm{IoD}<1$. In the context of counting statistics, the IoD is commonly referred to as the Fano factor~\cite{PhysRev.72.26}. The Fano factor plays a central role in mesoscopic physics, where it is used to characterize shot noise in electronic conductors~\cite{BLANTER20001}. In quantum optics, it distinguishes among super-Poissonian, Poissonian, and sub-Poissonian photon statistics~\cite{Fox2006QuantumOptics, PhysRevA.87.013847}. In particular, sub-Poissonian statistics ($\mathrm{IoD} < 1$) is a hallmark of non-classical light emitted by quantum sources such as single atoms~\cite{PhysRevLett.51.384} and quantum dots~\cite{Wang2019SubPoissonian}.

\begin{figure}[t]
    \centering
    \includegraphics[width=.85\linewidth]{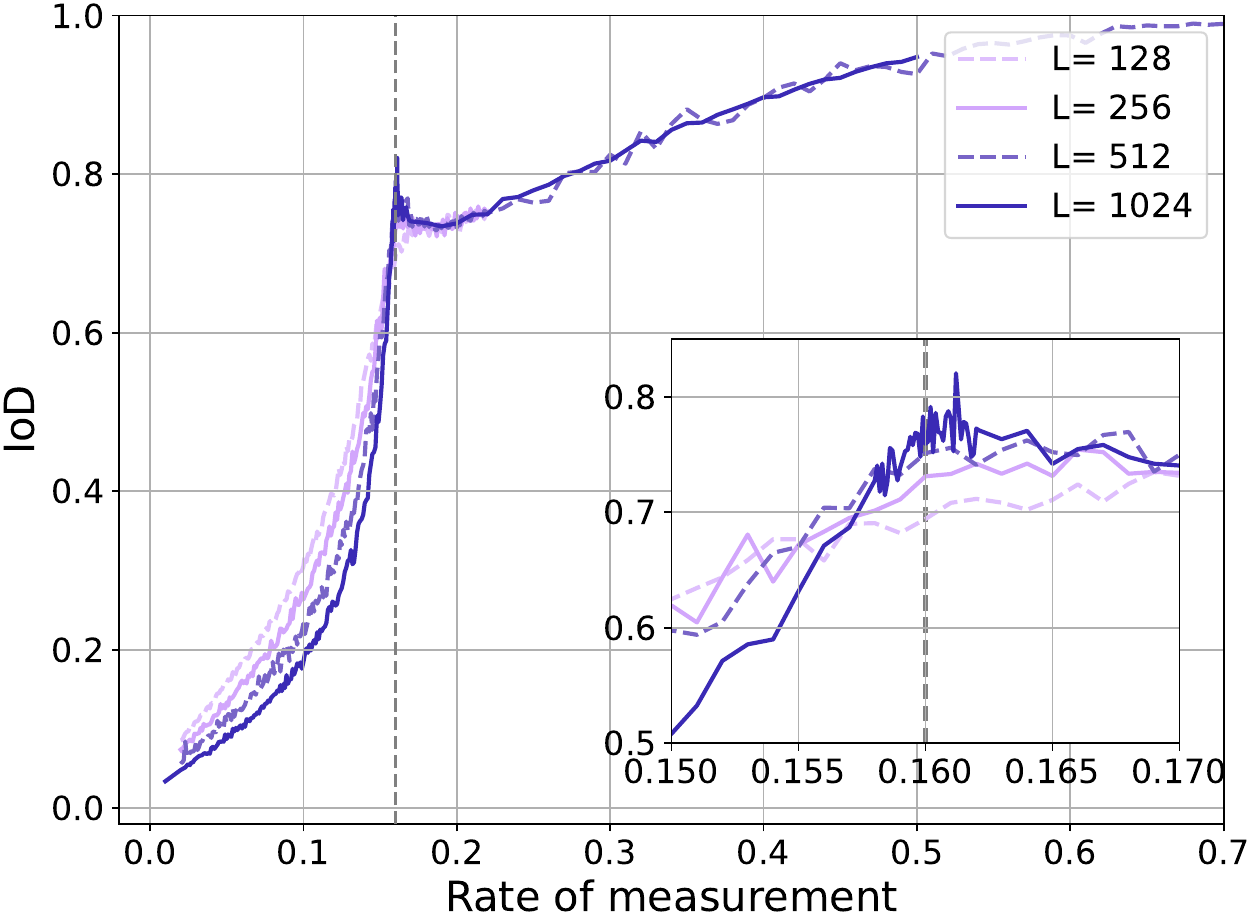}
    \caption{Index of dispersion (IoD) for varying rate of measurement. The two distinct phases are clearly observed in the IoD curves. The inset shows that the discontinuity in the curves occurs near the critical point, $p_c = 0.16$ (dashed vertical line).}
    \label{fig:curve_of_dispersion}
\end{figure}

As shown in the Fig.~\ref{fig:curve_of_dispersion}, we find that the IoD exhibits qualitatively distinct behavior across the two entanglement phases. In the volume-law regime, the IoD curves vary with different system sizes at the same measurement rate. Whereas in the area-law regime, the curves with different system sizes converge to a single curve. As the measurement rate increases, the IoD approaches unity, indicating that the entanglement entropy distribution becomes approximately Poissonian. The inset of Fig.~\ref{fig:curve_of_dispersion} shows the discontinuity between the volume-law regime and area-law regime, which occurs at the critical point, suggesting the possible use of the IoD to detect the phase transition.

We also highlight that computing the IoD requires no additional data beyond that needed to obtain the mean and variance, making it a simple yet effective diagnostic of the entanglement phase.

\begin{figure}[t]
    \centering
    \includegraphics[width=.85\linewidth]{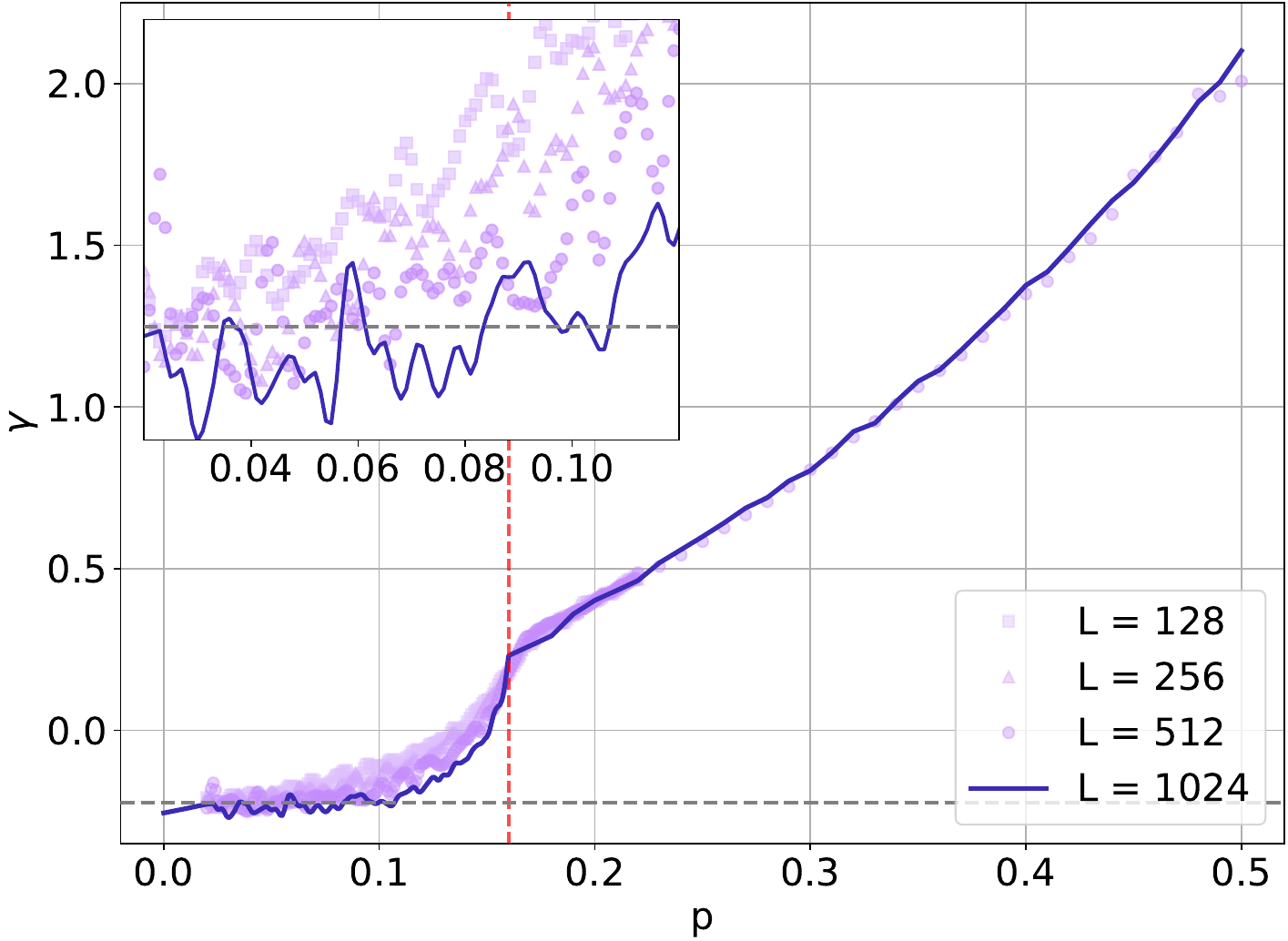}
    \caption{Skewness curves of the entanglement entropy distribution for various system sizes. Horizontal (gray) dashed line corresponds to the value of skewness predicted from the DPRE model. Vertical (red) dashed line represents critical probability, $p_c$. The inset shows data in the volume-law regime, where the skewness is distributed around the predicted value from the DPRE model, $\gamma = -0.224$, independent of the system sizes.}
    \label{fig:skew_curve}
\end{figure}
\subsection{Skewness}
While the IoD can efficiently indicate the critical behavior of MIPT using only the mean and variance of the entanglement distribution, these up to second-moment quantities do not capture the detailed shape of the distribution, for example, symmetry around the mean value. To explore a more detailed distribution of the entanglement entropy, we utilize the skewness, the lowest-order moment sensitive to the asymmetry of a given distribution. It is used in various areas of physics, including fluid dynamics~\cite{PhysRevFluids.6.104608}, plasma physics~\cite{PhysRevLett.108.265001}, and statistical physics~\cite{Wampler2022SkewnessKurtosis, PhysRevE.90.062402}. Especially in the context of statistical physics, higher-order moments of fluctuating quantities can be used to distinguish the universality class of the system~\cite{PhysRevE.90.062402}.

The skewness of the entanglement entropy distribution is defined as
\begin{equation}
    \gamma(S) = \frac{\langle(S-\langle S\rangle)^{3}\rangle}{(\mathrm{Var}[S])^{3/2}}.
\end{equation}
As a dimensionless and normalized quantity, the skewness is invariant under linear rescaling and shifts of $S$, making it insensitive to overall scale changes or mean shifts. This property is particularly advantageous in the present context, as it implies that additional finite-size normalization is not required when comparing different system sizes. Consequently, the skewness serves as a scale-free probe of asymmetry in the entanglement entropy distribution, enabling us to detect qualitative changes in its structure across the measurement-induced phase transition.

\begin{figure}[t]
    \centering
    \includegraphics[width=.8\linewidth]{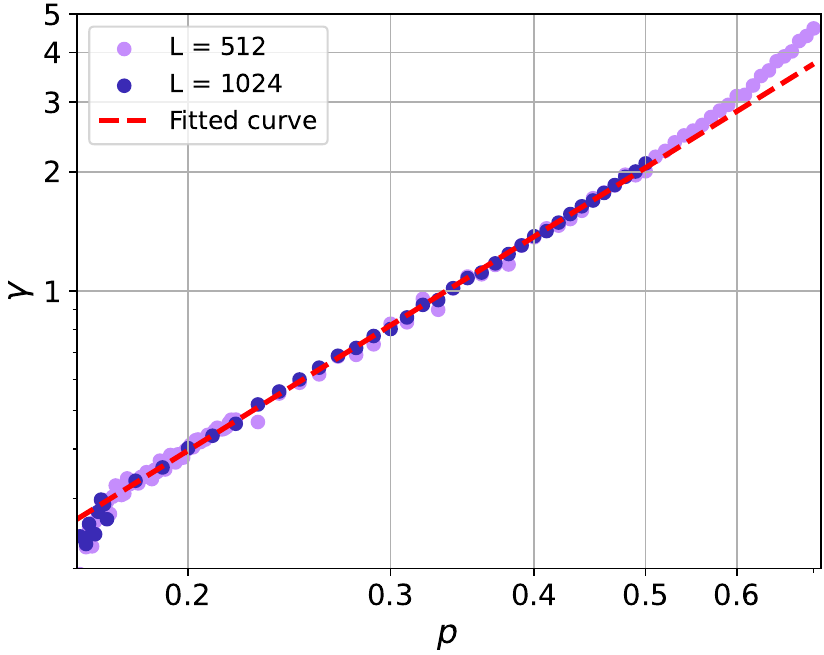}
    \caption{Log-log plot of skewness in the area-law regime. Skewness in the area-law regime closely follows a single curve, $\gamma \propto p^{1.79}$ which is indicated by a dashed red line.}
    \label{fig:area_skew}
\end{figure}

Figure~\ref{fig:skew_curve} shows skewness curves for different system sizes. In the volume law regime, despite the wide range of system sizes, the skewness remains at a constant value of $\gamma(S_{\mathrm{Vol}}) = -0.224$ regardless of the system sizes. This constant value results from a previously known mapping between a hybrid quantum circuit in the volume-law regime and a statistical-mechanics model~\cite{PRXQuantum.4.010331}. We defer the details on this matter to the latter part of this work. The invariance of skewness in the volume-law regime distinguishes it from other moments, such as the mean, variance, or IoD. On the other side of the phase transition, the skewness curve gradually increases in the area-law phase. This behavior becomes evident in a log-scaled plot as shown in Fig.~\ref{fig:area_skew}, revealing a power-law scaling with $p$, $\gamma(S) = 7.10 p^{1.79}$.

\begin{figure}[t]
    \centering
    \includegraphics[width=.8\linewidth]{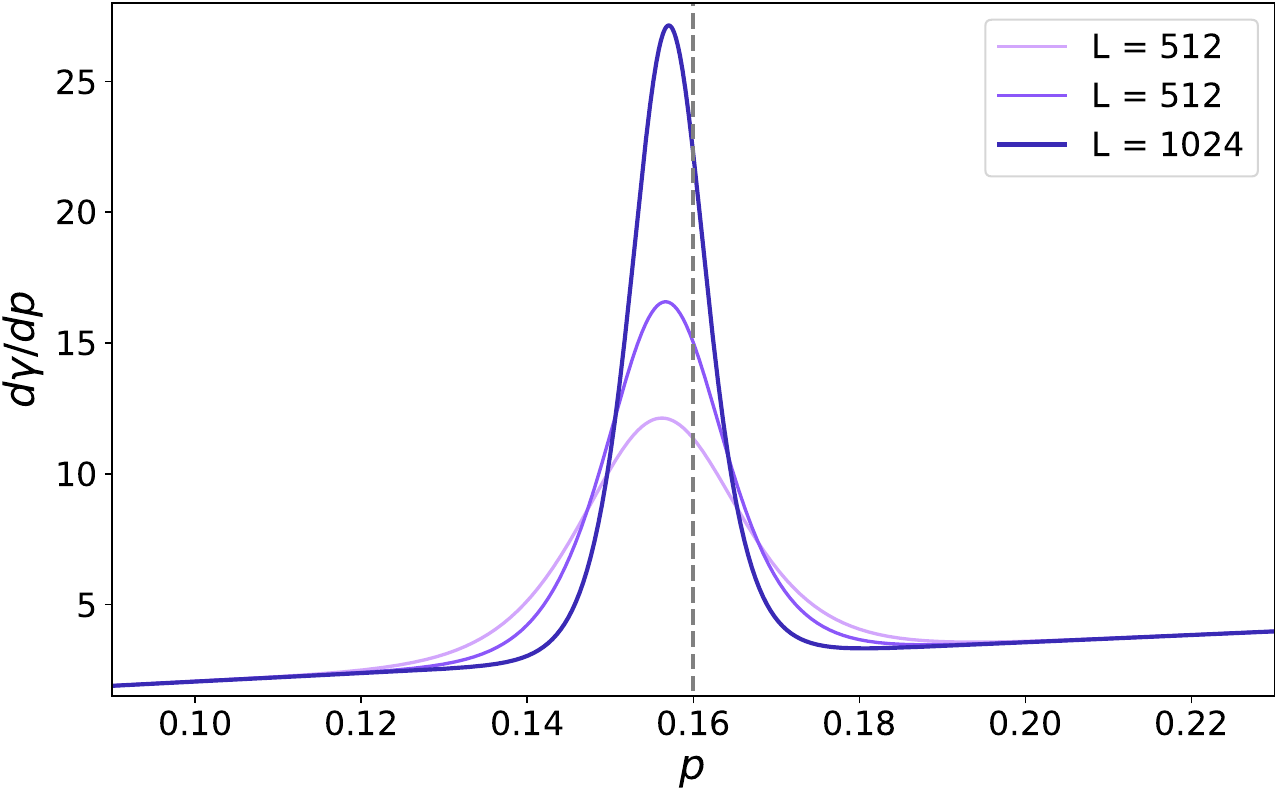}   
    \caption{Curves of numerical differentiation of the auxiliary function, Eq.~\eqref{eq:axillary_function}. The peak forms near the critical point, $p_c = 0.16$ (gray dashed vertical line), which narrows with increasing system sizes.}
    \label{fig:curve_fit}
\end{figure}
Between the two scaling phases, a sudden increase in skewness is observed, indicating significant deformation of the entanglement entropy distribution at the transition point. To address the point at which the change in skewness of distributions becomes the most significant, we employ a smooth auxiliary function to locate the local maximum. The choice of auxiliary function should be taken such that it describes the transition between the distinct behaviors of skewness in the two scaling phases near the criticality as follows:
\begin{equation}
    \gamma(S) = \begin{cases}
    -0.224, & \text{if } p \ll p_c \\
   7.10 p^{1.79}, & \text{if } p \gg p_c.
    \end{cases}
\end{equation}

Taking into account the power law scaling in the area-law regime, a smooth function was fitted to the skewness data for $L = 512$ and $1024$ to estimate the point of the maximal change in skewness. We take the following form of the auxiliary function,
\begin{equation}\label{eq:axillary_function}
    \gamma(p) = 7.10p^{1.79} + b_0 [\mathrm{tanh}(b_1(p-b_2))-1],
\end{equation}
where $b_{0,1,2}$ are constants obtained by minimizing mean-squared distances between the data and the curve for each system size. The first term captures the asymptotic power-law scaling in the area-law phase, whereas the latter term is introduced to model a crossover near the critical point. The hyperbolic tangent function provides a smooth interpolation between two asymptotic regimes and is therefore well-suited to describe a localized but continuous change in the slope of the skewness curve. In this manner, the function introduced in Eq.~\eqref{eq:axillary_function} serves as a phenomenological tool to robustly identify the point of maximal deformation in the entanglement entropy distribution. 
In Fig.~\ref{fig:curve_fit}, we present fits of the auxiliary function to our simulation results, which agree well with the data. 

With this choice of auxiliary function, the maximal points were found to be $p_{\mathrm{max}} = 0.158 \pm 0.009$ for $L = 1024$ and $p_{\mathrm{max}} = 0.155 \pm 0.012$ for $L = 512$. These values are close to the known critical value, $p_c = 0.16$, supporting our intuition that near the critical point, where the phase transition occurs, there is a maximal deformation in the shape of the distribution, which is well captured by the skewness. The errors were estimated using the full width at half maximum (FWHM) of the numerical differentials of the fitted curves. The FWHM is greater for $L=512$ than for $L=1024$, giving a less accurate estimate. This result indicates that higher-order moments, such as skewness, can be used to identify the MIPT transition point, which cannot be obtained from lower-order moments, such as the mean or the variance.

\section{Effective model for entanglement distribution}
\subsection{Volume law phase}

Recently, Li \textit{et al}.~\cite{PRXQuantum.4.010331} demonstrated an exact mapping between hybrid quantum circuits and a directed polymer in a random environment (DPRE). Using a replica approach, the entanglement entropy of hybrid circuits can be expressed in terms of the free energy cost of entanglement domain walls \cite{jian_measurement_induced_2020, PhysRevB.101.104301} which follows the same probability distribution as the free energy of the corresponding directed polymer~\cite{PRXQuantum.4.010331}. Specifically, the entanglement entropy is described by the following form
\begin{equation}\label{eq:DPRE}
    S \approx F =s_0L_A - s_1 (L_A)^{1/3}\xi,
\end{equation}
where $L_A$ is the subsystem size, which is taken to be $L/2$ in this work. $s_{0}$ and $s_1$ are non-universal constants, and $\xi$ is a random variable drawn from the Tracy-Widom (TW) distribution from the Gaussian unitary ensemble (GUE)~\cite{Tracy1994, kpz_cocktail}.
The periodic boundary conditions imposed in our circuit geometry select the GUE universality class for the associated directed polymer~\cite{PRXQuantum.4.010331,Johansson2000}.

\begin{figure}[t]
    \centering
    \includegraphics[width=.85\linewidth]{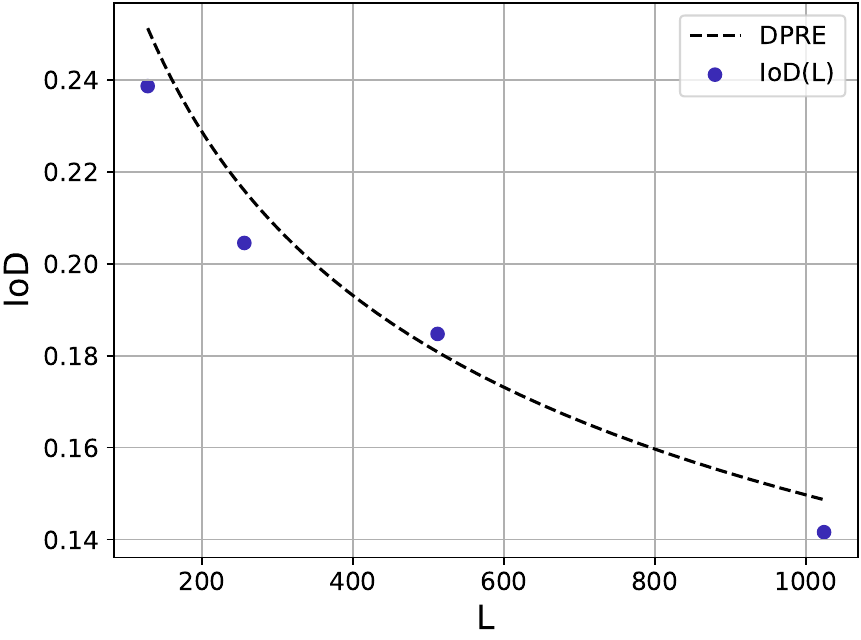}
    \caption{IoD as a function of system size $L$ for a fixed rate of measurement, $p=0.08$. It can be seen that the IoD data follows the trend predicted from the DPRE model.}
    \label{fig:IoD_DPRE}
\end{figure}

From the above equation, we can compute the IoD and skewness predicted from the DPRE model. Calculating the IoD reveals its dependence on system size, as shown in Fig.~\ref{fig:IoD_DPRE}:
\begin{equation}\label{eq:IoD_pred}
    \mathrm{IoD} = \frac{s_{1}^2 L_{A}^{1/3}\mathrm{Var}[\xi] }{s_0 L_{A}^{2/3} - s_1 \langle\xi\rangle},
\end{equation}
where $\langle\xi\rangle=-1.771$ and $\mathrm{Var}[\xi] = 0.8132$ are the average and the variance of the TW distribution, respectively~\cite{8160003}. For skewness, it can be easily  shown that the skewness of the entanglement entropy distribution coincides with that of the TW distribution:
\begin{align}
    \gamma(S) &= -\frac{s_{1}^{3}L_A\langle (\xi - \langle \xi\rangle)^{3}\rangle}{(s_{1}^{2}\mathrm{Var}[\xi])^{3/2}} \\
    &= -\gamma(\xi),
\end{align}
where $\gamma(\xi) = 0.224$~\cite{8160003}, in agreement with our numerical results in the volume-law regime. Therefore, it provides independent numerical confirmation of the DPRE description established in Ref.~\cite{PRXQuantum.4.010331} for the monitored quantum circuits with periodic boundary conditions.

\subsection{Area-law phase}

To describe the entanglement entropy in the area-law, we introduce the following stochastic model by assuming a simple structure for the entanglement entropy and its change after saturation. We consider the change in entanglement entropy over a single time step:
\begin{align}
    \Delta S_t &= \sigma_U - \sigma_{M}\\
    S_{t+1} &= S_t +\Delta S_t,
\end{align}
where $\sigma_{U(M)}$ is a random variable representing the effect of unitary (measurement) operations, which generally increase (decrease) the entanglement entropy. For simplicity, we assume that $\sigma_U$ follows a Gaussian distribution with mean $\mu = 0.58$ and variance of $\nu = 0.80$, both of which are determined numerically from simulations.

Within the stabilizer formalism, the entanglement entropy can be interpreted as the number of Bell pairs shared between two subsystems \cite{Ent_stab}. To model the effect of measurements, we therefore consider the probability that a measurement destroys a Bell pair, as well as the probability that measuring one qubit of the pair destroys the associated correlation. The probability that at least one qubit of a given Bell pair is destroyed, $q$, is
\begin{equation}
    q \equiv \frac{\binom{L}{N} -\binom{L-2}{N}}{\binom{L}{N}} ,
\end{equation}
which yields $q = 1 -(1-p)^2 + O(L^{-1})$ in the large $L$ limit. By assuming that each Bell pair has an equal probability of being destroyed, the average number of Bell pairs destroyed by measurements is given by $qS$. To calculate the variance of the number of Bell pairs destroyed, we must consider cases in which the destroyed qubits belong to the same pair. The probability that two distinct pairs are destroyed, $r$, is given by:
\begin{equation}
    r = \frac{\binom{L}{N}-2\binom{L-2}{N}+\binom{L-4}{N}}{\binom{L}{N}},
\end{equation}
and the variance is given by $q(1-q)S +(r-q^2)S(S-1)$, where $r-q^2$ is the covariance of the event that two distinct pairs are destroyed. $r-q^2$ can be calculated algebraically and in the large $L$ limit, $r-q^2 \xrightarrow{L \gg 1} -\frac{4p(1-p)^3}{L} + O(L^{-2})$. However, this form does not accurately describe the numerical data, as the correlation is strongly dependent on $q$, and we replaced $r-q^2$ with $c(q)$, a quadratic function of $q$ whose coefficients are optimized from our numerical data for each $p$.

Combining these ingredients, the conditional average and variance of $\Delta S$, $a(S)$ and $b(S)$ are
\begin{align}
    a(S) \equiv \langle(\Delta S|S)\rangle&= \mu - qS \\
    b(S) \equiv \mathrm{Var}(\Delta S |S) &= \nu + q(1-q)S +c(q) S(S-1).
\end{align}
These moments define a Fokker-Planck equation for the probability distribution $P(S,t)$,
\begin{align}
    \partial_t P(S,t) &=-\partial_S [ aP(S,t) ] + \frac{1}{2} \partial^{2}_{S}[bP(S,t)]\\
    &= -\partial_SJ(S,t)
\end{align}
where the second line is the continuity equation with probability current
\begin{equation}
    J(S,t) = a(S)P(S,t)-\frac{1}{2} \partial_S [b(S)P(S,t)].
\end{equation}
In the stationary state, $\partial_t P(S)=0$, so that $\partial_S J(S)=0$ and hence the current is independent of $S$. Imposing the zero-flux boundary condition at the boundaries of the allowed domain of $S$, ($0 \leq S \leq L/2$), we obtain $J(S)=0$, which yields
\begin{equation}
    a(S)P(S)-\frac{1}{2} \frac{d}{dS} [b(S)P(S)] = 0.
\end{equation}
Solving the equation gives the stationary distribution $P(S)$~\cite{Risken1989},
\begin{align} 
    P(S) &\propto \exp(\int  -\frac{b'}{b}+\frac{2a}{b} dS) \\
    &\propto \frac{1}{b}\exp(\int\frac{2(\mu -qS)}{\nu + \alpha S + c(q) S(S-1)} dS ),
    \label{eq:FP-sol}
\end{align}
where $b' = \frac{db}{dS}$ and $\alpha = q(1-q)$. Since the covariance $c(q)$ is small, in general, we treat it perturbatively and perform a Maclaurin expansion of the integrand up to second order in $c(q)$ (see Appendix~\ref{app:perturb} for more details).

We emphasize that this derivation of the Fokker-Planck equation is a coarse-grained stochastic description, not a fundamental equation of motion. Therefore, the equation neglects the microscopic details of the underlying circuit structure and dynamics. However, this approach allows a low-dimensional description to capture the entanglement entropy distribution in the area-law phase.

\subsection{Numerical simulation}
\begin{figure}[t]
    \centering
    \includegraphics[width=1\linewidth]{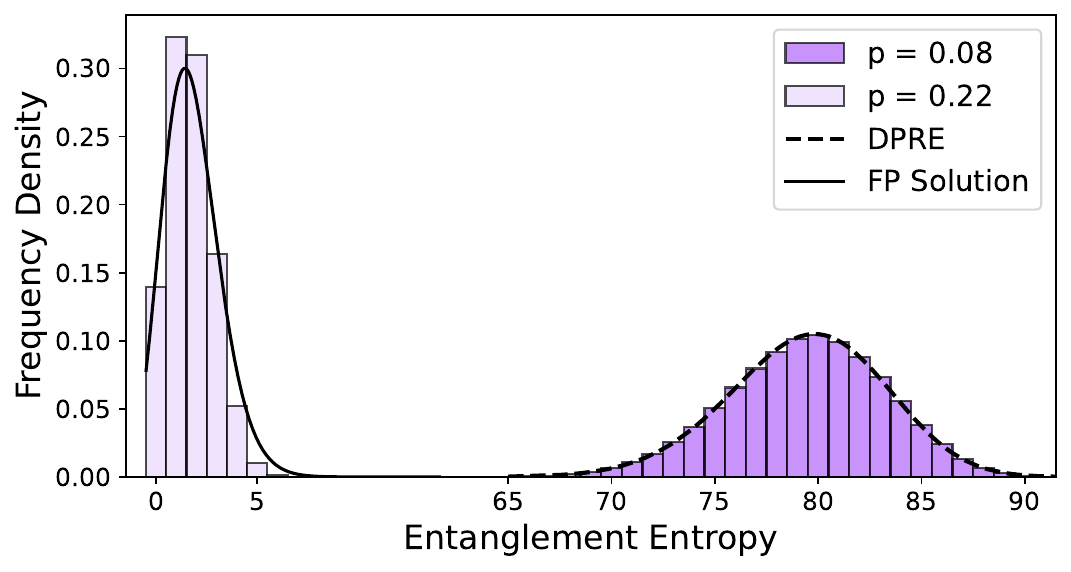}
    \caption{Entanglement entropy distributions in the volume law regime (right) and area-law regime (left). Distributions predicted by the DPRE model (dashed line) and the Fokker-Planck equation (solid line) capture the shape of the distribution. The entanglement entropy scale on the horizontal axis has been shifted so that the two distributions are visible together.}
    \label{fig:L512_histograms}
\end{figure}

We obtain the theoretical distributions in the volume-law regime by fixing the non-universal constants $s_0$ and $s_1$. From the Eq.~\eqref{eq:DPRE}, they can be found simultaneously from the $\langle S\rangle$ and $\mathrm{Var}[S]$ for a given $p$:
\begin{align}\label{eq:s1s0}
    s_1 &=\sqrt{\frac{\mathrm{Var}[S]}{\mathrm{Var}[\xi]L_{A}^{2/3}}} \\
    s_0 &= \langle S \rangle L_{A}^{-1} + s_1 \langle \xi\rangle L_{A}^{-2/3}.
\end{align} 

To obtain the solution of the Fokker-Planck equation, the parameters $\mu$, $\nu$, and the function $c(q)$ must be specified. The mean and variance of the entanglement entropy change induced by a single unitary layer, $\mu$ and $\nu$, are extracted numerically by tracking $\Delta S$ under one unitary layer, which we find to be $\mu=0.58$ and $\nu=0.80$. The covariance correction $c(q)$ is then determined by fitting the numerical data to a quadratic function of $q$, which gives $c(q) = 0.3285 q^2 -0.4536q+ 0.1239$. 

In Fig.~\ref{fig:L512_histograms}, we show a comparison between the histogram of our data and theoretical predictions as discussed above. The DPRE model accurately captures the distribution, whereas the solution of the Fokker-Planck equation captures it with moderate accuracy. 


\begin{figure}[t]
    \centering
    \includegraphics[width=.8\linewidth]{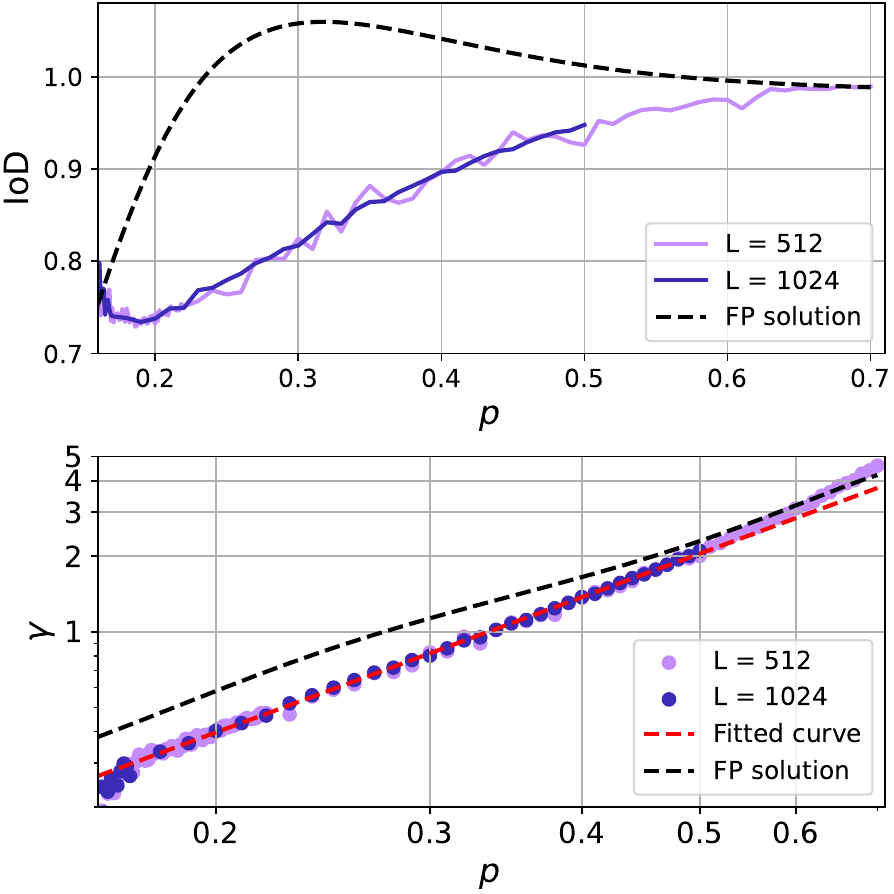}
    \caption{Comparison of IoD (top) and skewness (bottom) curves between the solution of the Fokker-Planck (FP) equations and the numerical data in the area-law regime.}
    \label{fig:FP_comparison}
\end{figure}
In Fig.~\ref{fig:FP_comparison}, we compute the IoD and skewness from the predicted distributions in the area-law regime. In terms of IoD, there is a large gap between the IoD predicted by the model and the simulated data, which closes in the large-$p$ limit. The skewness curve of the predicted distribution follows a similar trend to that of the fitted curve and the data. 

\begin{figure}[t]
    \centering
    \includegraphics[width=0.8\linewidth]{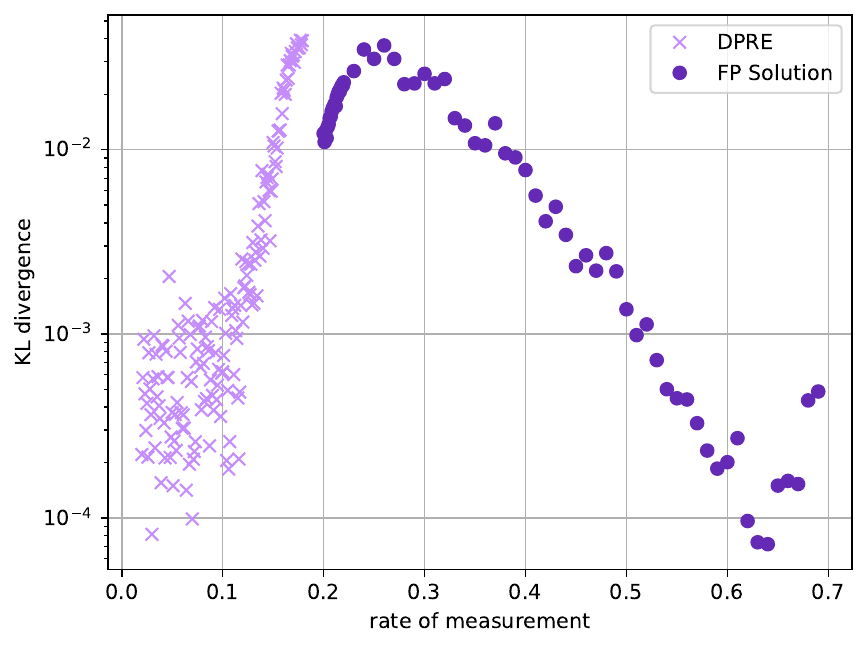}
    \caption{Kullback-Leibler divergence between the simulation data and the effective models derived from DPRE and the Fokker-Planck equation.}
    \label{fig:KL_log}
\end{figure}
Finally, to assess how accurately the predicted distributions describe the numerical data across a wide range of measurement rates, we compute the Kullback–Leibler (KL) divergence~\cite{Kullback1951} between the model prediction, $Q(S)$, and the simulated distributions, $P(S)$, where the KL divergence is defined as
\begin{equation}
    D_{\mathrm{KL}}(P||Q) = \sum_{s\in S} P(s) \log \frac{P(s)}{Q(s)}.
\end{equation}
Here, the sum over $s$ runs over all realizations of entanglement entropy trajectories with non-zero values of $P(S)$. The KL-divergence vanishes if and only if the two distributions are the same. Figure~\ref{fig:KL_log} shows that the TW distribution derived from the DPRE mapping and the solution of the Fokker-Planck equation provide an accurate description of the entanglement entropy distributions in the volume-law and the area-law phases, respectively, whereas both models become inaccurate as the measurement rate gets closer to criticality.

\section{Conclusion}
We have investigated the distribution of entanglement entropy in hybrid quantum circuits by examining higher-order moments beyond the mean. To this end, we have introduced the IoD and the skewness as second- and third-order probes of the entanglement entropy distribution, capturing the MIPT. Interestingly, while the variance alone might not be a reliable diagnostic of the MIPT, the IoD, which captures the relative width of the distribution, exhibits distinct behaviors across the volume-law and area-law phases and thus clearly indicates the critical point. We have also demonstrated that the skewness serves as a robust and scale-free diagnostic of the MIPT. Specifically, the constant value of skewness in the volume-law phase is in quantitative agreement with the TW distribution predicted from the DPRE mapping~\cite{PRXQuantum.4.010331}, providing independent numerical support for this theoretical description.

We have further explored effective descriptions of the entanglement entropy distribution in both phases. In the volume-law regime, the DPRE framework accurately reproduces the distribution and its higher moments. In the area-law regime, we have proposed a minimal coarse-grained stochastic model based on a Bell-pair picture within the stabilizer formalism. This model successfully captures the qualitative features of the distribution deep in the area-law phase, while revealing the limitations of a coarse-grained description near the critical regime.

Our results highlight that higher-order moments of the entanglement entropy distribution can provide a powerful methodology for characterizing the MIPT that cannot be accessed through the mean alone. Beyond the present setting, the methodology developed here can be applied to a broad class of stochastic quantum systems, in which fluctuations play a central role. This work thus opens a route toward a more complete statistical characterization of entanglement dynamics in monitored quantum systems.

\begin{acknowledgements}
    This work was supported by the National Research Foundation of Korea (NRF) grant funded by the Korea government (MSIT) (RS-2024-00413957 and RS-2024-00438415), the Institute of Information \& Communications Technology Planning \& Evaluation (IITP) grant funded by the Korea government (MSIT) (IITP-2026-RS-2020-II201606, IITP-2026-RS-2024-00437191, and RS-2025-02219034), and the Institute of Applied Physics at Seoul National University.  H.K. is supported by the KIAS Individual Grant No. CG085302 at Korea Institute for Advanced Study.
\end{acknowledgements}

\begin{appendix}
\section{Perturbative expansion of $P(S)$} \label{app:perturb}
We provide a Maclaurin expansion of the stationary distribution $P(S)$ of the Fokker-Planck equation in Eq.~\eqref{eq:FP-sol} as follows~\cite{Risken1989, BenderOrszag1999}:
\begin{equation}
P(S) \propto \frac{1}{b}P_0 P_1 P_2,
\end{equation}
where $P_0, P_1$ and $P_2$ collect the contributions of order $[c(q)]^0, [c(q)]^1$, and $[c(q)]^2$. Each term can be expressed as 
\begin{align}
    P_0 &= x^{\gamma}e^{-\frac{2S}{1-q}} \\
    P_1  &= x^{d_1 c(q)} \exp [-c(q) \{-d_0 x^{-1} + d_2 x  + \frac{d_3}{2} x^2  \}] \\
    P_2 &= x^{e_1 [c(q)]^2} \\
    & \quad \times \exp [-[c(q)]^2 \{-e_0 x^{-1} + e_2 x + \frac{e_3 }{2}x^2 +\frac{e_4 }{3}x^3 \}] ,
\end{align}
where $x =(\nu+\alpha S)$ and $\gamma = \frac{2}{\alpha}(\frac{\nu}{1-q} + \mu)$. The coefficients $d_i$ and $e_i$ are determined by expanding in $c(q)$ and then integrating over $S$: 
\begin{align}
    d_0 &= 2\left[\mu\left(\frac{\nu}{\alpha}\right) + (\mu + q) \left(\frac{\nu}{\alpha}\right)^2 -q\left(\frac{\nu}{\alpha}\right)^3\right]\\
    d_1 &= 2\left[\frac{3q\nu}{\alpha^3} - \frac{2\nu(\mu + q)}{\alpha^2} - \frac{\mu}{\alpha} \right]\\
    d_2 &= 2\left[\frac{\mu + q}{\alpha^2} - 3\frac{q\nu}{\alpha^3}\right]\\
    d_3 &= 2\frac{q}{\alpha^3},
\end{align}
and
\begin{align}
    e_0 &= -2\left[ \frac{\mu}{\alpha}\nu +  \frac{2\mu+q}{\alpha^2}\nu^2 +  \frac{\mu+2q}{\alpha^3}\nu^3 +  \frac{q}{\alpha^4}\nu^4 \right]\\
    e_1 &= 2\left[\frac{\mu}{\alpha}+2\left(\frac{2\mu+q}{\alpha^2}\right)\nu +3\left(\frac{\mu+2q}{\alpha^3}\right)\nu^2 +4\left(\frac{q}{\alpha^4}\right)\nu^3 \right]\\
    e_2 &= -2\left[ \frac{2\mu+q}{\alpha^2}+3\left( \frac{\mu+2q}{\alpha^3}\right)\nu    +6\left(\frac{q}{\alpha^4}\right)\nu^2\right]\\
    e_3 &= 2\left[ \frac{\mu+2q}{\alpha^3}     +4\left(\frac{q}{\alpha^4}\right)\nu\right]\\
    e_4 &= \frac{-2q}{\alpha^4}.
\end{align}

\end{appendix}
\bibliography{ref}

\end{document}